\documentclass[aps,prl,reprint,superscriptaddress,graphicx]{revtex4-1} 
\usepackage{amsmath}
\usepackage{graphicx}

\draft 

\begin{document}


\title{Epitaxial lift-off for solid-state cavity quantum electrodynamics}



\author{Lukas Greuter}
\affiliation{Department of Physics, University of Basel, Klingelbergstrasse 82, Basel 4056, Switzerland}

\author{Daniel Najer}
\affiliation{Department of Physics, University of Basel, Klingelbergstrasse 82, Basel 4056, Switzerland}

\author{Andreas V. Kuhlmann}
\affiliation{Department of Physics, University of Basel, Klingelbergstrasse 82, Basel 4056, Switzerland}

\author{Sascha Valentin}
\affiliation{Lehrstuhl f\"ur Angewandte Festk\"orperphysik,Ruhr-Universit\"at Bochum, D-44780 Bochum, Germany}

\author{Arne Ludwig}
\affiliation{Lehrstuhl f\"ur Angewandte Festk\"orperphysik,Ruhr-Universit\"at Bochum, D-44780 Bochum, Germany}

\author{Andreas D. Wieck}
\affiliation{Lehrstuhl f\"ur Angewandte Festk\"orperphysik,Ruhr-Universit\"at Bochum, D-44780 Bochum, Germany}

\author{Sebastian Starosielec}
\affiliation{Department of Physics, University of Basel, Klingelbergstrasse 82, Basel 4056, Switzerland}
\author{Richard J. Warburton}
\affiliation{Department of Physics, University of Basel, Klingelbergstrasse 82, Basel 4056, Switzerland}


\date{\today}

\begin{abstract}
We present a new approach to incorporate self-assembled quantum dots into a Fabry-P\'{e}rot-like microcavity. Thereby a 3$\lambda$/4 GaAs layer containing quantum dots is epitaxially removed and attached by van der Waals bonding to one of the microcavity mirrors. We reach a finesse as high as 4,100 with this configuration limited by the reflectivity of the dielectric mirrors and not by scattering at the semiconductor - mirror interface, demonstrating that the epitaxial lift-off procedure is a promising procedure for cavity quantum electrodynamics in the solid state. As a first step in this direction, we demonstrate a clear cavity-quantum dot interaction in the weak coupling regime with a Purcell factor in the order of 3. Estimations of the coupling strength via the Purcell factor suggests that we are close to the strong coupling regime.

\end{abstract}


\maketitle 


The interaction of optically active semiconducting nanostructures such as quantum dots (QDs) with light can be massively increased by placing the emitter into a microcavity, thereby allowing a study of cavity quantum electrodynamics (CQED) in the solid state.  A measure of the cavity-emitter interaction is the cooperativity parameter, $C = 2g^2/\kappa\gamma$, which puts all involved rates in context: the emitter-cavity coupling rate $g$, the cavity photon decay rate $\kappa$ and the emitter decay rate $\gamma$. If $g\ll\kappa,\gamma$, the system is in the weak coupling regime and an emitted photon is irreversibly lost before it can be reabsorbed. However, the increased photon density of states associated with the cavity mode results in an accelerated spontaneous emission when the QD and cavity are in resonance~\cite{Purcell1946}. This enhancement of the spontaneous emission increases the quantum efficiency of single photon sources~\cite{Vuckovic2003,Strauf2007}, an important feature for many applications in quantum information~\cite{Shields2007}.
If $g\gg\kappa,\gamma$, the system is in the strong coupling regime and an energy quantum is coherently and reversibly exchanged between emitter and cavity mode resulting in new eigenstates, polartions i.e.\ superpositions of cavity photon and emitter excitation. Strong coupling is the prerequisite for the realization of a single photon transistor~\cite{Birnbaum2005} and enables QD-QD coupling with potential applications in quantum information~\cite{Imamoglu1999}. Both regimes were already observed with self-assembled QDs in micropillars~\cite{Gerard1998, Reithmaier2004, Lermer2012} and photonic crystals~\cite{Chang2006, Yoshie2004, Englund2012, Hennessy2007}. Alternative experiments with QDs coupled to fully tunable microcavities showed a clear Purcell enhancement~\cite{Barbour2011, Muller2009, Miguel-Sanchez2013} and even allowed the observation of strong coupling~\cite{Greuter}. 

The threshold to observe a finite splitting in polariton energy is given by $4g > \left|\kappa-\gamma\right|$ from the Jaynes-Cummings Hamiltonian that describes the coupled system~\cite{Auffeves2010}. While $\gamma$ is an intrinsic property of the emitter, $g$ and $\kappa$ can be tailored by choosing an appropriate cavity design. The cavity decay rate $\kappa$ is limited by the reflectivity of the mirrors and characterized by the quality factor $Q = \omega/\kappa$, where $\omega$ is the resonance angular frequency of the cavity. 
The coupling strength $g$ is given by $\hbar g = \mu_{12}E_{\rm vac}$, where $\mu_{12}$ is the emitter's dipole moment and the vacuum field $E_{\rm vac}\propto 1/\sqrt{V_0}$ scales inversely with the cavity mode volume $V_0$. Thus, efforts to achieve a strong QD-cavity coupling seek to decrease the mode volume at high cavity $Q$-factors. 

Generally, for QDs coupled to micropillars or photonic crystal cavity modes, the benefit of a small mode volume comes at the cost of the $Q$-factor. Furthermore, spectral and spatial tunability remains limited in these cavities. Other approaches with tunable cavity designs~\cite{Hunger2010, Dufferwiel2014, Greuter2014} so far incorporate the InGaAs QDs in a heterostructure which also contains a semiconductor distributed Bragg reflector (DBR) consisting of several pairs of AlGaAs/GaAs. These two materials have the same lattice constant but a different refractive index ($n_{\rm AlGaAs} = 3.009$, $n_{\rm GaAs} = 3.54$). There are three issues here. First, this material combination (equal lattice constant but significantly different refractive index) is unique to GaAs, unfortunately limiting DBR-based CQED applications to self-assembled InGaAs QDs. Second, the relatively low refractive index contrast results in a high penetration depth of the cavity field into the mirror thus enlarging the mode volume. Finally, semiconductor 'supermirrors' are essentially impossible to fabricate: growth of more than, say, 40 pairs is extremely time consuming.

We present here a best-of-both-worlds approach for CQED: it combines the benefits of a solid state emitter with a low-loss high-reflectivity dielectric DBR. The tantalizing possibility is to embed a fast, robust solid-state emitter in a low mode volume microcavity formed using dielectric supermirrors. We remove epitaxially a thin GaAs layer containing InGaAs QDs and bond the layer via van der Waals (VdW) forces to a dielectric DBR. This forms one of the end mirrors in the Fabry-P\'{e}rot-like tunable microcavity. The InGaAs QD is a suitable candidate, partly due to its relative short recombination time (i.e. large oscillator strength~\cite{Dalgarno2008}). Dielectric mirrors enable an ultrahigh finesse (up to $10^6$~\cite{Rempe1992}) with also a small penetration depth. We note that the approach shown here enhances greatly the flexibility of cavity experiments: each part of the cavity (bottom mirror, top mirror, optically active layer) can be individually designed and processed and then combined to create an optimized CQED system. 

We demonstrate a successful epitaxial lift-off (ELO) and subsequent attachment to a dielectric DBR by van der Waals bonding. The mirror is then integrated into the tunable cavity design and we show that the finesse of the cavity remains high despite the presence of the new GaAs-DBR interface. Furthermore, we show a weak coupling of a single QD to the microcavity mode as revealed by a reduction of the lifetime when the QD is in resonance with the  cavity. In fact an estimation of the coupling $g$ implies that our system is close to the strong coupling regime and we state that by minor improvements the observation  of the typical anticrossing is within reach in this system.


\begin{figure}
\includegraphics[width=85mm]{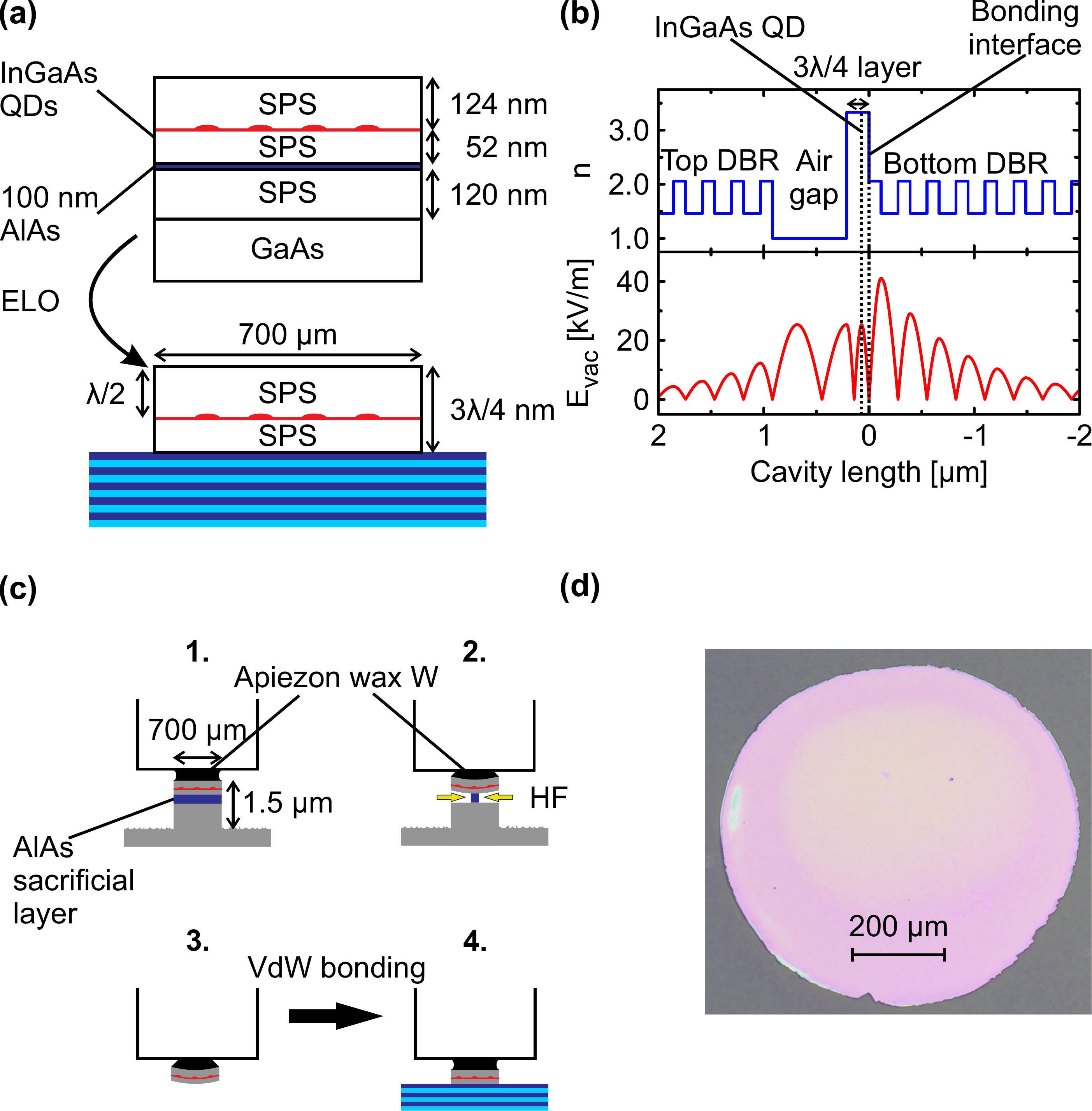}
\caption{Epitaxial lift-off (ELO) procedure for cavity QED with QDs. (a) Sample structure before (top) and after (bottom) the ELO process. (b) 1D transfer matrix method simulation of the microcavity design with the bonded ELO layer. The design is chosen such that a field node is located at the bonding interface. (c) The fabrication consists mainly of  the following steps: 1.\ Mesa-etching and attachment to an acid-resistant teflon stamp; 2.\ etching of the sacrificial AlAs layer in 10$\%$ HF until the GaAs substrate falls away; and 3.\ and bonding to a dielectric DBR. (d) Optical microscope image of a 3$\lambda$/4 thick GaAs layer bonded to a dielectric DBR after ultrasonic bath cleaning. }
\label{fig:ELO}
\end{figure}

In the present proof of concept experiment, we bond a 3$\lambda$/4 epitaxial layer that embeds the QDs onto a Ta$_2$O$_5$/SiO$_2$ DBR ending with the high refractive index material (Ta$_2$O$_5$). This serves as one mirror in the tunable cavity, while the other consists of a DBR ending with a Ta$_2$O$_5$ layer as well. A 1D transfer matrix method simulation of the vacuum field for this particular cavity design is shown in figure~\ref{fig:ELO}b. By design, an electric field node is located at the GaAs epilayer-mirror interface with the hope that fabrication imperfections may have only a limited effect on the finesse. This is a conservative approach: the penetration depth into the bottom mirror could be further reduced by bonding a $\lambda$ layer on a DBR ending with SiO$_2$, but with the drawback of a field antinode at the interface. 

The sample before (after) the epitaxial lift-off is shown in figure~\ref{fig:ELO}a, top (bottom). The heterostructure is grown by molecular beam epitaxy (MBE) on a GaAs substrate followed by a $225$ nm thick GaAs layer that includes a $120$ nm thick short period supperlattice (SPS). On top of the SPS a $100$ nm thick AlAs layer is grown as sacrificial layer followed by the ELO layer with a total thickness of 3$\lambda/4$ for a design wavelength of $\lambda = 940$ nm. The ELO layer contains the InGaAs QDs grown at a distance $\lambda/2$ from the surface such that they are located at an antinode of the vacuum field in order to maximize the coupling to the cavity (figure~\ref{fig:ELO}b). The QDs are surrounded by another two SPSs resulting in an average refractive index of the complete ELO layer $n = 3.332$.

Figure~\ref{fig:ELO}c shows the process of the epitaxial lift-off procedure and the subsequent van der Waals bonding. For the separation of the ELO-film, we first deposit a small piece of Apiezon wax on the sample and heat it to 125 $^\circ$C for 1 hour. The melted wax defines a round structure with diameter of $\approx 700~\mu$m for etching a mesa with a solution consisting of sulfuric acid (H$_2$SO$_4$), hydrogen peroxide (H$_2$O$_2$) and deionized water (H$_2$O) with a volume ratio of 1:8:120, commonly known as piranha-solution.  We first etch $\approx 1.5~\mu$m with the piranha solution such that the AlAs sacrificial layer is exposed for the subsequent etching with hydrofluoric acid (HF). The piranha-etched sample with the wax is then reheated to $70 - 80~^\circ$C and attached to a homemade teflon stamp before immersed into the $10~\%$ HF solution ( step 1.\ and 2.\ in figure~\ref{fig:ELO}c). The epitaxial lift-off is based on the high selectivity ($10^8 : 1$) on etching AlAs in GaAs in a $10~\%$ HF-solution. During the etching process with HF, the stress induced by the surface tension of the wax bows the epitaxial layer and ensures an open etching channel~\cite{Yablonovitch1990,Demeester1993}. 

After the AlAs sacrificial layer is completely etched, the substrate falls away and the ELO-film stays attached to the teflon stamp (3.\ in figure~\ref{fig:ELO}c). The HF-solution is then highly diluted ($\ll 0.001\% $) by rinsing with deionized water before the new host substrate is immersed into the liquid. The host substrate consists of silica coated with a distributed Bragg reflector (DBR) (design reflectivity $99.98 \%$) consisting of alternating $\lambda/4$ layers Ta$_2$O$_5$ ($n = 2.06$) and  SiO$_2$ ($n = 1.46$) ending with Ta$_2$O$_5$. The ELO-film remains immersed in the solution throughout the exchange of the substrates and the subsequent bonding is conducted completely in DI-water. This provides a very clean environment and hence minimizes contamination with particles between the two surfaces~\cite{Yablonovitch1990}. The weight of the stamp results in pressure on the order of a few $\rm N/mm^2$ on the sample during the bonding process (step 4.\ in figure~\ref{fig:ELO}c).  After the highly diluted HF solution is poured away, the sample is dried for 24 h. Ideally, any remaining water film at the bonding interfaces evaporates and the ELO-film is pulled down by surface tension such that close range (VdW) forces bond the layer to the substrate. Experimentally however, a small gap between the interfaces can emerge during the bonding process as shown below. The bonded sample is then detached from the stamp by removing the wax with trichloroethylene (TCE). The resulting bottom mirror structure after bonding is shown in figure~\ref{fig:ELO}b. The bonding strength of VdW-bonding is sufficient high. Evidence for this is the optical microscope image in figure~\ref{fig:ELO}d, which shows an intact ELO-film bonded to a DBR mirror even after the immersion in an ultrasonic bath.


\begin{figure}
\includegraphics[width=85mm]{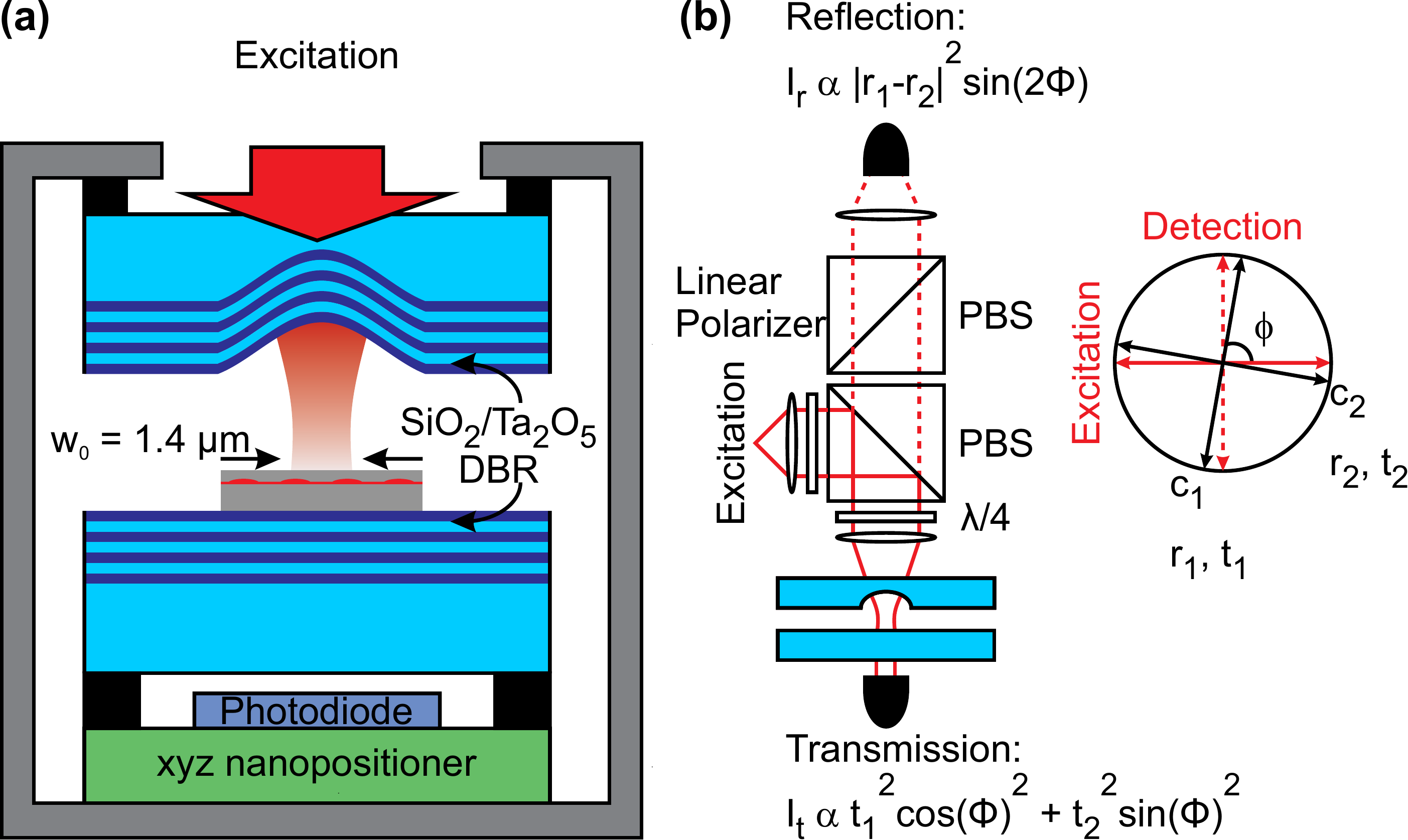}
\caption{(a) Tunable microcavity setup. The ELO-layer is bonded on the dielectric DBR and the entire sample is mounted on an xyz-positioner stage that can be positioned with respect to the concave top mirror. (b) Cross-polarized darkfield detection scheme realized by two polarizing beam splitters. The polarization axes of the cavity mode are only slightly misaligned with respect to the axes of the microscope head. 
\label{fig:setup}
}
\end{figure}

\begin{figure}
\includegraphics[width=85mm]{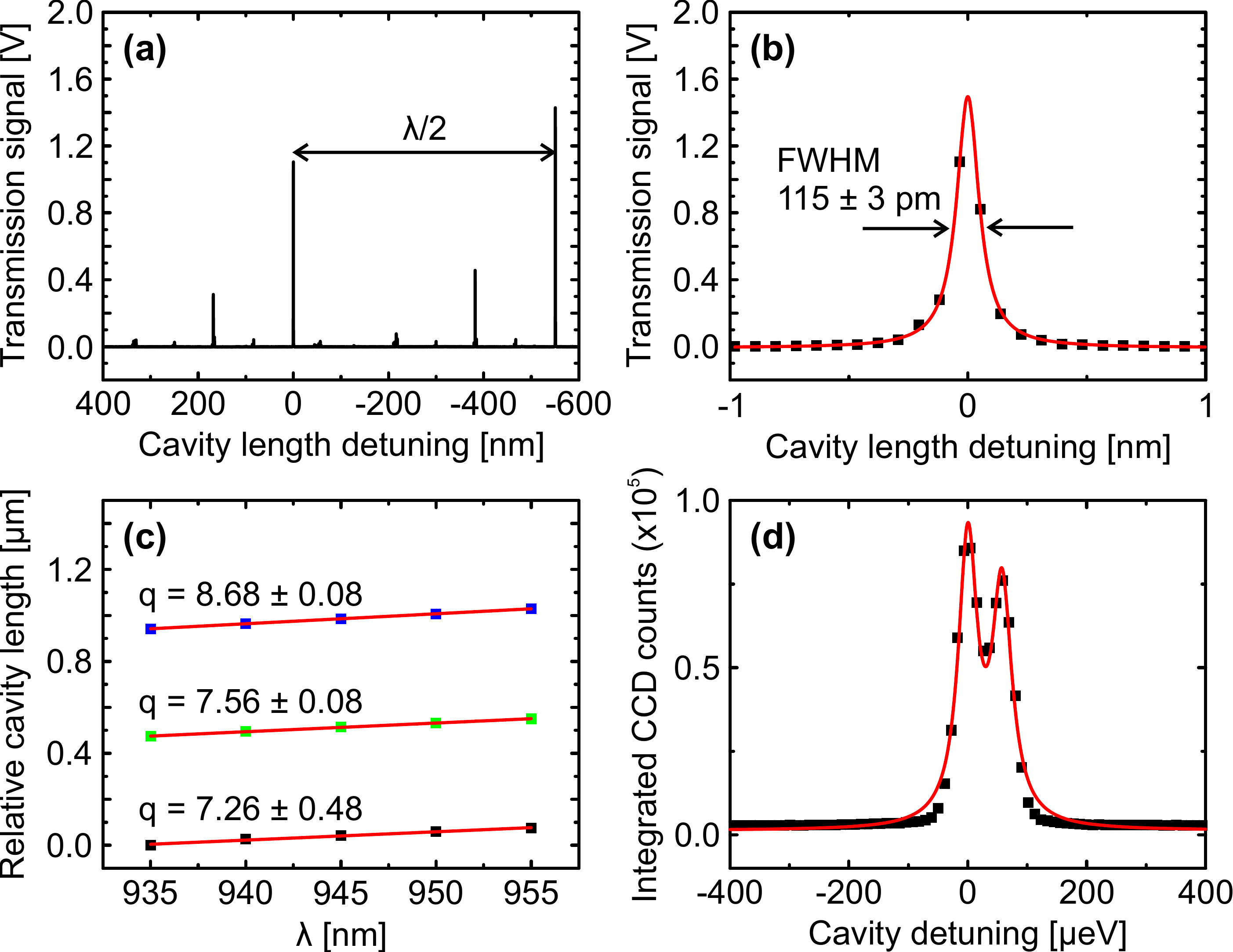}
\caption{Performance of the cavity with an embedded ELO layer. (a) Transmission measurement revealing the fundamental and higher order cavity resonances. (b) A single transmission peak with full width half maximum (FWHM) of $\delta d = 115~\pm~3~\rm pm$ results in a Finesse of $4,100~\pm~100$. (c) Determination of the smallest accessible mode index by varying the excitation wavelength. (d) A cross-polarized detection of the cavity mode in reflection reveals a mode splitting not observed in transmission (see text).
\label{fig:cavityperformance}
}
\end{figure}

The DBR with the bonded ELO-film serves as an end mirror in the fully tunable plane-concave microcavity setup shown in figure~\ref{fig:setup}a and described in detail in~\cite{Greuter2014}. Here, the top mirror consists of a concave mirror with radius of curvature $R$ of $13~\mu\rm m$, fabricated by CO$_2$-laser ablation and subsequent coating with a Ta$_2$O$_5$/SiO$_2$ DBR, exactly the same DBR as the substrate for ELO bonding. Spectral and spatial tunability is realized by mounting the bottom mirror on an xyz-piezo positioner such that it can be displaced in three dimensions with respect to the top mirror. We study the performance of the microcavity including the ELO-bonded bottom mirror at $4~\rm K$ in a He bath cryostat. 

For cavity excitation and detection, we interrogate the system with a coherent cw laser (linewidth $1~\rm MHz$) with a cross-polarized detection scheme realized by incorporating two polarizing beam splitters (PBS) in the microscope head at room temperature (figure~\ref{fig:setup}b)~\cite{Kuhlmann2013}. The cavity is excited with a fixed linear polarization, while only light orthogonally polarized to the excitation is detected. An additional Si-photodetector mounted directly underneath the bottom mirror facilitates cavity transmission measurements.

Figure~\ref{fig:cavityperformance}a shows a measurement of the cavity transmission signal as a function of cavity-length detuning for a fixed probe laser beam wavelength $\lambda = 940~\rm nm$. We identify two fundamental cavity modes at physical distance of $\lambda/2$, accompanied by higher order modes. The structure of the cavity mode is described by Hermite-Gaussian $\rm TEM_{qnm}$ modes, where the transversal mode splitting is determined by the radius of curvature of the top mirror. Figure~\ref{fig:cavityperformance}b shows one cavity resonance with Lorentzian lineshape and a full width at half maximum linewidth (FWHM) of $\delta d = 115~\pm~3~\rm pm$. We identify the finesse in the spatial domain to be $F = \lambda/(2\delta d) = 4,100~\pm~100$. The absolute mode index $q = 2\delta d/\delta\lambda$ is determined by varying the probe wavelength for the first 3 available cavity modes (figure:~\ref{fig:cavityperformance}c). We reach a minimum mode index of $7.26~\pm~0.48$ which translates to an effective cavity length of $l = 3.4~\pm~0.2~\mu m$. This length together with the measured finesse yields a quality factor of $Q = 2lF/\lambda = 30,000$ corresponding to a cavity linewidth of $44~\rm\mu eV$. 
When measuring in reflection (figure~\ref{fig:cavityperformance}d), we observe a fundamental mode splitting of $57.65~\rm \mu eV$. The two modes reveal linewidths of $38.53$ and $40.29~\rm \mu eV$ respectively ($Q$-factors: $34,200$ and $32,700$), agreeing well with the linewidths expected from finesse and cavity length measurements in transmission.

The fundamental mode splitting of the microcavity is indicated as $c_1$ and $c_2$ in figure~\ref{fig:setup}b. The two modes are linearly polarized and we speculate that the splitting arises in the ELO-layer. We note that the linear polarization axes of the $c_1$, $c_2$ modes coincide with the crystallographic axes of the epitaxial lift-off layer. The birefringence may be due to strain-induced anisotropy~\cite{Ziel1977} and was also shown to determine the polarization properties of semiconductor vertical-cavity surface-emitting lasers (VCELs)~\cite{JansenvanDoorn1996}.

Figure~\ref{fig:setup}b shows the alignment of our cross-polarized detection scheme with respect to the linear polarized cavity modes with an angle $\phi$ close to $\pi/2$. The two cavity modes can be characterized by the detuning dependent reflection (transmission) coefficients $r_1, r_2$ ($t_1, t_2$) that obey the relation $r^2+t^2=1$.  We excite the cavity with an electric field amplitude $E_0$ along the excitation axis and detect orthogonally to it. The total signal that is projected on the detection axis is composed of the two electric amplitudes $E_1$ and $E_2$ that originate from the two cavity modes $c_1$ and $c_2$:
\begin{subequations}\label{eq:efielddetection}
\begin{eqnarray}
E_1 & = & \frac{r_1}{2}E_0\sin{2\phi}, \\
E_2 & = & -\frac{r_2}{2}E_0\sin{2\phi}.
\end{eqnarray}
\end{subequations}
\noindent We detect an intensity in reflection $I_r$ = $\left|E_1+E_2\right|^2$:
\begin{equation}
I_r = \frac{I_0}{4}\left|r_1-r_2\right|^2\sin^2{2\phi}.
\label{eq:detection}
\end{equation}
\noindent For the transmission intensity $I_t$ the signal depends solely on the alignment of the excitation with the cavity axis and we can derive:
\begin{equation}\label{eq:transmission}
I_t = t_1^2\cos^2{\phi}+t_2^2\sin^2{\phi}.
\end{equation}
\noindent These two equations suggest that for an angle $\phi$ close to 90$^\circ$ only the $c_1$ mode is efficiently detected in transmission, while in reflection the signal is proportional to the contrast of the detuning-dependent reflection coefficients $r_1$ and $r_2$.

In the present design, the refractive indices satisfy $n_{\rm Ta_2O_5}^2 \approx n_{\rm ELO}n_{\rm SiO_2}$, close to the condition for an anti-reflection (AR) coating which results in a penetration depth of $6.70~\rm\mu m$ into the bottom mirror and a minimal total cavity length of $7.32~\rm\mu m$  when simulated with a 1D transfer matrix method. This is significantly larger than the $3.4~\rm\mu m$ estimated from the absolute mode index in figure~\ref{fig:cavityperformance}c. We explain this discrepancy between theory and experiment by an imperfect bonding of the ELO layer to the bottom mirror. A simulation of a $22~\rm nm$ thick gap with refractive index $n = 1$ (equivalently a $17~\rm nm$ thick H$_2$O-film) between the ELO layer and the bottom mirror supports this assumption. Such a configuration breaks the condition of the AR coating and the penetration depth is reduced to 2.06 $\mu$m. Together with an air gap of $\approx 0.5~\rm\mu m$ (due to imperfect parallelism of the two mirrors) we calculate a total cavity length of $3.00~\rm\mu m$, in accordance with the measurement.


 \begin{figure}
\includegraphics[width=85mm]{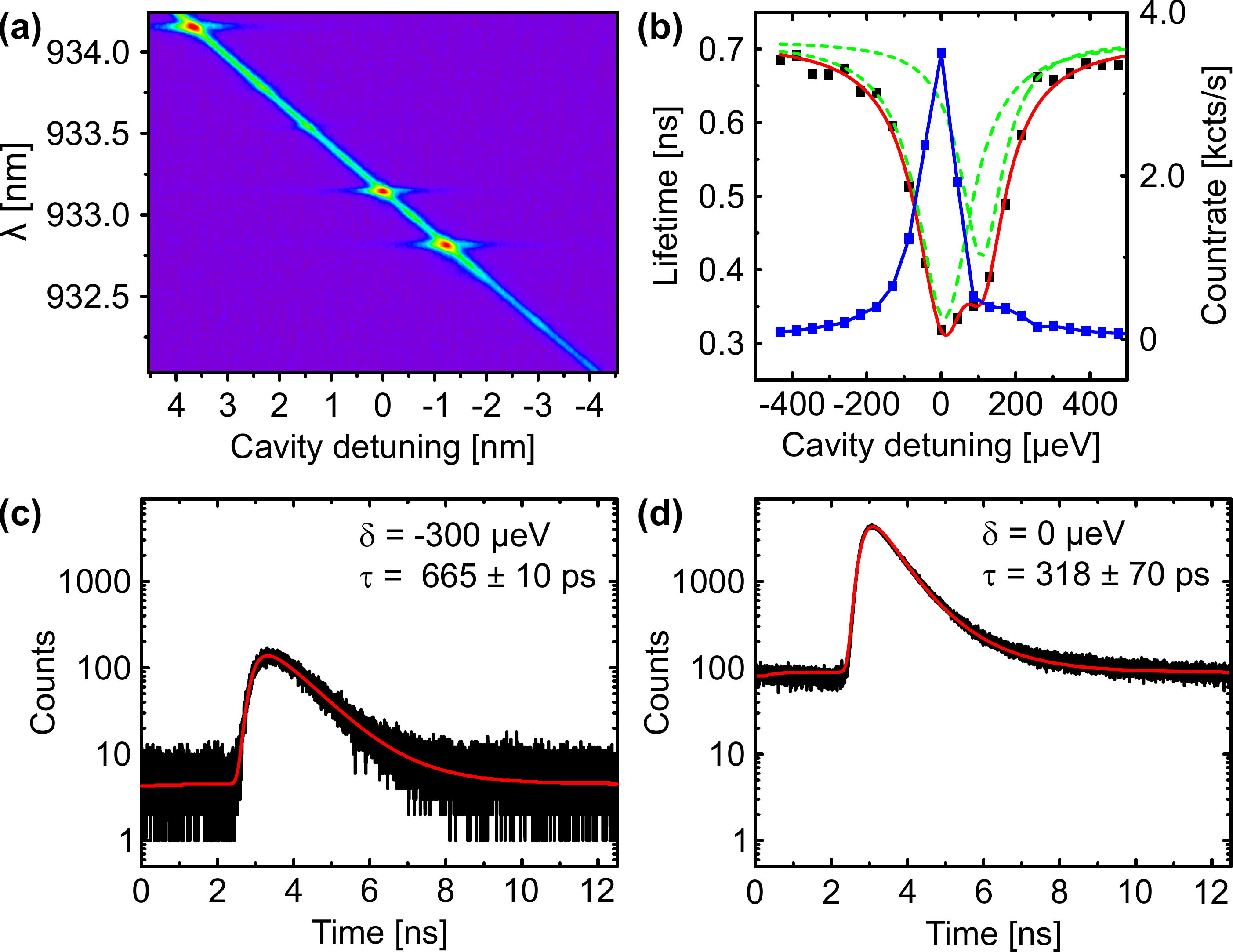}
\caption{Quantum dot in a tunable microcavity. (a) PL spectrum as a function of cavity tuning. Distinct bright points signify the emission from single QDs. (b) Lifetime measurements (black dots) of the QD with $\lambda_{QD} = 933.18$ nm as a function of cavity-dot detuning. A clear reduction of the lifetime is observed on resonance owing to the Purcell effect. The blue dots show the simultaneously recorded total counts. The red curve is a fit to equation~\ref{eq:Purcell}, where the green dashed lines indicate the two Lorentzians of the fit. (c),(d) Lifetime measurement of the cavity 300~$\mu$eV detuned and on resonance. Single exponential fits taking into account the internal response function reveal lifetimes of 665 ps detuned and 318 ps on resonance. 
\label{fig:QDincavity}
}
\end{figure}

We demonstrate weak coupling of a single InGaAs QD to our tunable microcavity by means of photoluminescence (PL). The cavity-QD system is nonresonantly excited ($\lambda = 830$ nm) and the emitted signal is analysed by a CCD-based spectrometer with a spectral resolution of 40 $\mu$eV. The cavity resonance is tuned by applying a voltage on the z-piezo, which acts on the cavity length. Figure~\ref{fig:QDincavity}a shows the spectrum as a function of cavity detuning. We detect an emission into the cavity mode for all values of $z$. Earlier crosscorrelation measurements~\cite{Winger2009, Kaniber2008} interpreted this background as a hybridization of the higher QD states with the neighboring wetting layer~\cite{Karrai2004}. The discrete bright spots at specific energies indicate the coupling of a single QD to the cavity mode. The QD transitions observed here do not show a fine-structure splitting, a characterization feature of the neutral exciton $X^0$. It is likely that the transitions observed here are charged excitons $X^{-1}$.

To verify that the enhanced PL at the cavity resonance represents more than spectral filtering of the QD, we study the cavity-QD dynamics in the time domain for a single QD with $\lambda = 933.14~\rm nm$. The cavity-QD system is excited by a $Q$-switched pulsed nonresonant laserdiode with repetition rate of $80~\rm MHz$ and a pulse width of $50~\rm ps$. The excitation pulse defines the start signal. The stop signal is provided by the detection of an emitted photon by an avalanche photodiode with a timing resolution of $340~\rm ps$ and a dark-count rate of $20$ counts per second. Figure~\ref{fig:QDincavity}c shows a lifetime measurement of the cavity-QD system at 300 $\mu$eV detuning. A single exponential fit taking into account the internal response function (IRF) reveals a lifetime of $665~\pm 10 ~\rm ps$. At zero detuning (figure~\ref{fig:QDincavity}d), the lifetime of the QD is reduced to $318~\pm 70~\rm ps$ implying an increased spontaneous emission rate. For comparison, the lifetime of five different QDs not coupled to the cavity mode were measured yielding an average lifetime of $805~\pm~150~\rm ps$.  

Figure~\ref{fig:QDincavity}b shows lifetime measurements for different cavity-QD detunings. Due to the cross-polarized detection, only one cavity mode is observed in PL (figure~\ref{fig:QDincavity}a). However, the QD still couples to both orthogonally polarized modes and we expect a lifetime reduction in both cases. Thus the simultanously recorded countrate (blue dots in figure~\ref{fig:QDincavity}b) reduces at a detuning where the lifetime still remains low. Consequently, we fit the lifetime data in figure~\ref{fig:cavityperformance}b with a double-Lorentzian:  	 
\begin{equation}
\frac{\gamma_{\rm cavity}}{\gamma_{\rm free}} = \frac{F_{\rm P1}\Delta_1^2}{4\delta_1^2+\Delta_1^2}+\frac{F_{\rm P2}\Delta_2^2}{4\delta_2^2+\Delta_2^2}+\alpha.
\label{eq:Purcell}
\end{equation}
\noindent According to the measurements performed without a cavity, we take a free space decay rate of $\gamma_{\rm free} = 1.25~\pm~0.23~\rm GHz$. $\gamma_{\rm cav}$ is the decay rate into the cavity mode. The first two terms in equation~\ref{eq:Purcell} describe the cavity-QD detuning-dependent relative decay rate according to the density of states in the two microcavity modes. The term $\alpha$ describes the relative decay rate into leaky modes of the cavity. $F_{\rm P1}$ ($F_{\rm P2}$) is the Purcell factor corresponding to the first (second) cavity mode, $\Delta_1$ and $\Delta_2$ are the two cavity mode linewidths and $\delta_1$ ($\delta_2$) is the cavity-QD detuning with respect to the first (second) cavity mode.

From the fit, we determine Purcell factors of $F_{\rm P1} = 1.27~\pm~0.04$ , $F_{\rm P2} = 0.79~\pm~0.04$ and $\alpha = 1.12~\pm~0.01$. The corresponding linewidths are $\Delta_1 = 121.83~\pm~5.8~\rm\mu eV$ and $\Delta_2 = 106.93~\pm~10.71~\rm\mu eV$ and the splitting between the modes is $100.14~\pm~5.11~\rm\mu eV$. The errors on the determination of the two Purcell factors $F_{\rm P1}$ and $F_{\rm P2}$ arise from the statistical error of $\gamma_{\rm free}$, while the errors in the widths arise from the double-Lorentzian fit. The linewidth values are significantly larger than those in figure~\ref{fig:cavityperformance}d, where the cavity was probed with a laser and a cross-polarized detection scheme. As an explanation for this finding we note that we integrated for 30 minutes for each point in figure~\ref{fig:QDincavity}b in order to achieve low noise decay curves. On these long timescales the cavity resonance drifts slightly since we do not actively stabilize the cavity length during measurement. This wandering of the cavity broadens the lineshape of the cavity-dot detuning dependent lifetime by a factor of $\approx 2.5$. Therefore we argue that the estimated Purcell factors represents the lower limit that is reached in the present setup (with a $\gamma_{\rm free}$ of 1.25 GHz). Correcting for the drift we evaluate $F_{\rm P1}' \approx 3.2$ and $F_{\rm P2}' \approx 2$. We note also that if the $c_1$, $c_2$ splitting could be eliminated, $F_{\rm p}$ would rise to $\sim 5$.

Once the Purcell factor is known an effective mode volume can be estimated by:
\begin{equation}
V_0 = \frac{3Q\left(\lambda/n\right)^3}{4\pi^2F_{\rm p}},
\label{eq:modevolume}
\end{equation}

\noindent with a $Q$-factor of $33,000$ and an averaged refractive index $n$ of the ELO-layer of $3.332$. A free space decay rate of $1.25~\rm GHz$ translates into a dipole moment $\mu_{12}$ of $1.2~\rm nm$ and we calculate the coupling $g$ with
\begin{equation}
g=\sqrt{\frac{\mu_{12}^2\omega}{2\epsilon_{0}n^2\hbar V_0}}.
\label{eq:coupling}
\end{equation}

\noindent We find $\hbar g = 11.75~\rm\mu eV$. 
We estimate a vacuum field $E_{\rm vac}$ at the location of the QDs via $\hbar g = \mu_{12}E_{\rm vac}$ to be $E_{\rm vac} = 1.0\times~10^4$ V/m. This is lower compared to the, $2.5~\times 10^4~\rm V/m$, calculated from the transfer matrix simulation. Imperfections at the bonding interface may shift the maximum of the vacuum field such that the location of the QD does not coincide with the electric field antinode. However, simulations with an airgap of 22 nm between the ELO layer and the bottom mirror suggest that this effect is negligible: the shorter penetration depth results in an increased vacuum field compensating for the vertical displacement of the antinode.

Since we do not have immediate access to the QD uncoupled to the cavity in our setup, the free space decay rate $\gamma_{\rm free}$ remains the uncertain factor for the Purcell factor estimation. However, the estimation of the coupling strength $g$ is stable with respect to $\gamma_{\rm free}$. This is due to the fact that in our analysis the Purcell factor scales with $1/\gamma_{\rm free}$ and hence $g$ $\propto 1/\sqrt{\gamma_{\rm free}}$, while simultaneously $g$ is proportional to $\mu_{12} \propto \sqrt{\gamma_{\rm free}}$ and the $\gamma_{\rm free}$ - dependency of $g$ cancels out. 

We point out that with a coupling rate $\hbar g \approx 12~\rm\mu eV$ and cavity rate $\hbar\kappa \approx 40~\rm\mu eV$ our cavity-QD dynamics are already close to the strong coupling regime $4g > \left|\kappa-\gamma \right|$. Moreover, our setup offers several possibilities to improve the cooperativity factor. Notably $\kappa$ can be significantly reduced by using 'supermirrors' with an ultrahigh reflectivity. Furthermore, as mentioned above, the current configuration at the interface of $3\lambda$/4 ELO/ $\lambda/4$ Ta$_2$O$_5$ / $\lambda/4$ SiO$_2$ results in a relatively high penetration depth and an unwanted reduction of the vacuum field strength. Now that a high quality ELO layer-DBR mirror is established, a $\lambda$-layer ELO can be bonded to a SiO$_2$-terminated DBR. Simulations suggest a reduction of the penetration depth from $6.70~\rm\mu m$ to $4.30~\rm\mu m$ and an increase of the vacuum field strength by factor 1.4 - already enough to observe a clear strong-coupling signature. To reduce the mode volume further, one possibility would be to use materials for the mirrors with a higher refractive index contrast.


In conclusion we have demonstrated a hybrid high-$Q$, low mode-volume tunable Fabry-P\'{e}rot microcavity consisting of a thin GaAs epilayer and dielectric mirrors. The fundamental requirements for cavity QED are met in this system: the finesse was high despite the new interface between dissimilar materials and the QDs remained optically active with low linewidths. Furthermore, we verified that the QD-cavity system operates in the weak coupling regime close to strong coupling. We argue that our epitaxial lift-off approach opens new possibilities for cavity QED in the solid state. 
  
\begin{acknowledgments}
We acknowledge financial support from SNF (project 200020-156637) and NCCR QSIT. 
\end{acknowledgments}




%

\end{document}